\documentclass[prb,twocolumn,superscriptaddress,10pt,floatfix,superscriptbibliography]{revtex4-2}
\setcitestyle{super}
\usepackage{amsmath}
\usepackage{amssymb}
\usepackage{graphicx}
\usepackage{bm}
\usepackage[top=2cm, bottom=3.0cm, left=2.0cm, right=2.0cm]{geometry}
\usepackage{mathrsfs}
\usepackage{subfigure}
\usepackage{booktabs}
\usepackage{float}
\usepackage{siunitx}
\usepackage{soul}
\usepackage[dvipsnames]{xcolor}
\usepackage[section]{placeins}
\usepackage[utf8]{inputenc}
\usepackage{ulem}

\def\be{\begin{equation}}
\def\ee{\end{equation}}
\def\bea{\begin{eqnarray}}

\def\bea{\end{eqnarray}}
\def\eea{\end{eqnarray}} 	
\def\bs{\begin{split}}
\def\es{\end{split}}
\def\ni{\noindent}

\def\mbf{\mathbf}

\def\bi{\begin{itemize}}
\def\ei{\end{itemize}}

\def\a{\alpha}
\def\b{\beta}

\def\o{\omega}

\def\f{\frac}
\def\tf{\tfrac}

\begin{document}

\title{\Large{Free energy barrier and thermal-quantum behavior \\ of sliding bilayer graphene}}
\author{Jean Paul Nery}
\email{nery.jeanpaul@gmail.com}
\affiliation{Dipartimento di Fisica, Universit\`a di Roma La Sapienza, I-00185 Roma, Italy}
\affiliation{Graphene Labs, Fondazione Istituto Italiano di Tecnologia, Via Morego, I-16163 Genova, Italy}
\affiliation{Nanomat group, QMAT research unit, and European Theoretical
Spectroscopy Facility,  Université de Liège, B5a allée du 6 août, 19,
B-4000 Liège, Belgium}
\author{Lorenzo Monacelli}
\affiliation{Dipartimento di Fisica, Universit\`a di Roma La Sapienza, I-00185 Roma, Italy}
\author{Francesco Mauri}
\email{francesco.mauri@uniroma1.it}
\affiliation{Dipartimento di Fisica, Universit\`a di Roma La Sapienza, I-00185 Roma, Italy}
\affiliation{Graphene Labs, Fondazione Istituto Italiano di Tecnologia, Via Morego, I-16163 Genova, Italy}
\vspace{10 mm}

\begin{abstract} 
In multilayer graphene, the stacking order of the layers plays a crucial role in the electronic properties and the manifestation of superconductivity.
By applying shear stress, it is possible to induce sliding between different layers, altering the stacking order. Here, focusing on bilayer graphene, we analyze how ionic fluctuations alter the free energy barrier between different stacking equilibria.
We calculate the free energy barrier through the state-of-the-art self-consistent harmonic approximation, which can be evaluated at unstable configurations.
We find that  above 100 K there is a large reduction of the barrier of more than 30\% due to thermal vibrations,
which significantly improves the agreement between previous first-principles theoretical work and experiments in a single graphite crystal.
As the temperature increases, the barrier remains nearly constant up to around 500 K, with a more pronounced decrease only at higher temperatures.
Our approach is general and paves the way for systematically accounting for thermal effects in free energy barriers of other macroscopic systems.
\end{abstract}

\maketitle

\section{Introduction}

There has been intense research on graphene since its discovery in 2004, and also on multilayer graphene systems. Among many other interesting properties, flat bands are present in multilayer rhombohedral (ABC-stacked) graphene (RG)\cite{Koshino2010}, and superconductivity has been observed in RG in 2021\cite{Zhou2021}. Several works recently studied, both experimentally and theoretically, under which conditions RG can form \cite{Yankowitz2014,Geisenhof2019,Bouhafs2021,Gao2020,Li2020,Kerelsky2021}, given that AB-stacked graphene is the more common occurring phase. 
In particular, some of us unveiled how the stacking of graphene layers changes under shear stress \cite{Nery2020}, and obtained that it produces many consecutive layers of RG. In fact, all samples with long-range RG that have been observed so far have been subject to some form of shear stress \cite{Lin2012,Henni2016,Henck2018,Yang2019,Shi2020,Bouhafs2021}.

To move graphene layers with respect to each other, they have to overcome a free energy barrier that separates different minima. When doing so, they experiment ``stick-slip" motion: they move gradually with increasing shear stress, followed by sudden rearrangements and decrease of the stress\cite{ZeLiu2012}. The maximum shear-stress to ``unlock" layers is correlated with the height of the barrier. Such barrier corresponds to the activation energy of a metastable state, where a graphene flake sitting on top of a large graphene layer is displaced relative to the equilibrium position. 

Superlubricity is a another phenomenon where the energy landscape corresponding to the relative displacement of layers is also central. In this mechanism, which has attracted a lot attention\citep{Liu2014,Sinclair2018}, particularly in carbon-based systems like multilayer graphene and graphite, incommensurate surfaces (i.e. atomic lattices without matching periodicity) lead to extremely low friction.
Yet another phenomena involving sliding surfaces is thermolubricity, which refers to the assistance from thermal excitations to overcome barriers. This term has been coined by some works that have studied the transition from stick-slip motion into stochastic fluctuations in friction at sufficiently high temperatures\cite{Baykara2018}. 
Despite the interest in sliding surfaces, there are no studies that account for both temperature and quantum effects of free-energy barriers in macroscopic systems.
A few works have studied quantum effects in free-energy barriers, including zero-point renormalization and tunneling, but of systems involving hydrogen molecules, including dissociative adsorption\cite{Mills1994} and diffusion\cite{Poma2012,McIntosh2013,Cendagorta2016}.

The variation of the potential along a periodic crystal surface, combined with the attraction between layers, 
prevents free sliding between them.
 This phenomenon has been extensively studied in various systems, like graphite\cite{Dienwiebel2004},  MoS$_2$\cite{Li2016}, silicon on gold\cite{Liu2015}, among others. 
When using a tip with the goal of making layers slide, normal forces need not be large. For pressures much lower than 1GPa, the alteration of the free-energy landscape is minimal (see SI of Ref.~\citenum{Nery2020}), so the relaxed barrier remains the dominant contribution that impedes sliding. In addition, some works on superlubricity have observed that friction has an extremely weak dependence with the load, attributed for example\cite{Dienwiebel2004} to a contact area that remains nearly constant in the force range applied. 
Here, we focus on calculating the height of the barrier, which should be approximately proportional to the force --given by the maximum slope of the free energy profile-- that is needed for a layer to move from one minima to another in commensurate systems.  
 
A quantity closely related to the free energy barrier is the shear frequency, corresponding to the mode in which layers move rigidly in-plane and out of phase around the equilibrium position. The free energy barrier separates two equivalent minima, so the barrier and shear frequency are expected to have a similar temperature dependence (a lower curvature at the minima corresponds to a lower shear frequency and a lower free energy, at least close to the minima). The value of the shear frequency in graphite is a relatively well-established quantity, with experimental values\cite{Lebedeva2017} oscillating between 42 - 45 cm$^{-1}$, using for example inelastic X-ray scattering\cite{Mohr2007}, neutron-coherent inelastic scattering\cite{Nicklow1972}, Raman spectra\cite{Tan2012,Hanfland1989}, and coherent phonon spectroscopy \cite{Boschetto2013}. In bilayer graphene, there are some experiments with values around 30 cm$^{-1}$; for example, 28 $\pm$ 3 cm$^{-1}$ in Refs.~\citenum{Amelinckx1965,Boschetto2013}, and 32 cm$^{-1}$ in Ref.~\citenum{Tan2012}. The temperature dependence of the shear modes, however, has been barely studied. Few experiments measure the temperature dependence of the shear mode, using  Raman scattering in folded 2+2 (4 layer) graphene\cite{Cong2014}, or femtosecond pump-probe spectroscopy in graphite\cite{Mishina2000}.

Another aspect of multilayer graphene and graphite that still needs further investigation and that serves as a consistency check on the adequacy of our approach is thermal expansion (TE). Ref.~\citenum{McQuade2021} measured it in bilayer and trilayer graphene for the first time. Ref.~\citenum{Mittal2021} uses the quasi-harmonic approximation (QHA), which underestimates the out-of-plane TE at low temperatures\cite{Bailey1970} and has a very rapid increase at high temperatures; in-plane, it also underestimates the TE at low temperatures\cite{Bailey1970} and overestimates it at high temperatures\cite{Morgan1972}. 
Much better out-of-plane results were obtained by using path integral molecular dynamics (PIMD)\cite{Herrero2020}, but while the TE flattens at 500 K in graphite, experimental values continue increasing\cite{Morgan1972,Marsden2018}. However, in-plane PIMD results of Ref.~\citenum{Herrero2020} vastly overestimate the TE.

Motivated by the aforementioned phenomena, especially by the possibility of facilitating sliding through temperature,
we focus on thermal effects on the free-energy barrier.
 This is a very challenging problem due to the difficulty of determining the free energy in a non-equilibrium position. 
Here we tackle it using the state-of-the-art stochastic self-consistent harmonic approximation (SSCHA)\cite{Errea2014,Monacelli2021}. First, in Sec.~\ref{sec:num_methods}, we briefly introduce SSCHA and the QHA, and explain how the barrier is determined. Then, in Sec.~\ref{sec:results}, we present our main results and compare to the height of the barrier with fixed atoms. Subsequently, using also the QHA, we calculate the temperature dependence of the shear frequency and compare to the limited experimental data available. Finally, we determine the in-plane and out-of-plane TE coefficient in bilayer graphene and graphite, and compare our results to previous theoretical and experimental work. In Sec.~\ref{sec:conclusions}, we summarize the conclusions of our work.
\vspace*{\fill}

\section{Theoretical and Computational Framework}
\label{sec:num_methods}

Here, we briefly describe the QHA, SSCHA, and a new SSCHA interpolation method used to determine the barrier. A more detailed description of SSCHA can be found in Refs.~\citenum{Errea2014} or \citenum{Monacelli2021}.

\subsection{Quasi-harmonic approximation (QHA)}
\label{sec:QHA}

In the QHA, at each value of the lattice parameters $\{a_i\}$, the free energy $F(T,\{a_i\})=U(\{a_i\})-TS(T,\{a_i\})$, where $U$ is the internal energy, $S$ the entropy and $T$ the temperature, is given by the standard harmonic expression. That is,

\be
\begin{split}
F(T,\{a_i\}) =  & U_\mathrm{lat}(\{a_i\}) + \f{1}{N_q} \sum_{\mathbf{q}\nu} \f{1}{2} \hbar \o_{\mathbf{q}\nu}(\{a_i\}) + \\
& + \f{1}{N_q}\sum_{\mathbf{q}\nu} k T \mathrm{ln} \left(1 - e^{-\hbar \o_{\mathbf{q}\nu}(\{a_i\})/k T}\right). \\
\end{split}
\label{eq:F_harmonic}
\ee

\ni where $U_\mathrm{lat}\{a_i\}$ is the internal energy of the lattice for a given value of the lattice parameters, $N_q$ the number of phonon wavevectors $\mbf{q}$, $\nu$ the mode, and $\o_{\mbf{q}\nu}$ the corresponding phonon frequency.

The value of the lattice parameters at a given temperature, $a_i(T)$, is determined by minimizing the free energy. Using an interatomic potential, the phonon frequencies at each value of the lattice parameters can be determined with little computational effort, and thus the QHA offers a quick method to determine the lattice expansion. Using first-principles calculations, phonons can be determined using density functional perturbation theory, which uses a primitive cell and is much faster than frozen phonon (finite differences) calculations in supercells. However, it does not account for the ``true anharmonicity''\cite{Allen2020}, present at fixed values of the parameters when varying temperature. For example, it gives an increasing as opposed to decreasing temperature dependence of the G mode in graphene, due to the negative TE of the lattice parameter. Also, if the TE is negative and the lattice parameter reduces below the classical lattice value, as in the case of monolayer graphene at about 400 K (see the Supplementary Information (SI), Fig.~S12), %Fig.~\ref{fig:SSCHA_QHA_a}),
 then an acoustic mode becomes unstable and the QHA is ill-defined. These issues are not present in SSCHA.

\subsection{Stochastic self-consistent harmonic approximation (SSCHA)}
\label{sec:SSCHA}

In general, the free-energy of a ionic Hamiltonian $H=T+V$ is given by

\be
F_H =  \mathrm{tr}(\rho_H H) + \f{1}{\b} \mathrm{tr}(\rho_H \mathrm{ln} \rho_H),
\ee

\ni where $\rho_H$ is the density matrix. In SSCHA, the density matrix is restricted to a trial harmonic Hamiltonian $\mathcal{H}=T+\mathcal{V}$, and the free energy is given by

\be
\mathcal{F}_H(\mathcal{H}) = \mathrm{tr}(\rho_\mathcal{H}H) + \f{1}{\b} \mathrm{tr}(\rho_\mathcal{H} \mathrm{ln} \rho_\mathcal{H}).
\ee

Due to a variational principle, $F_H \leq \mathcal{F}_H(\mathcal{H})$. The parameters of the harmonic Hamiltonian $\mathcal{H}$ include the centroids of the atoms \textcolor{black}{(their average positions)}, which correspond to the lattice parameters $a_i$ of the QHA, but it also includes the interatomic force constant matrix, which contains the information of all the phonon frequencies and accounts for the true anharmonicities. Thus, SSCHA has additional parameters, and is expected in general to give better results than QHA. Also, QHA does not satisfy a variational principle. It can be shown that

\be
\mathcal{F}_H(\mathcal{H})=F_\mathcal{H}+\int d\mathbf{R}[V(\mathbf{R})-\mathcal{V}(\mathbf{R})]\rho_\mathcal{H}(\mathbf{R}),
\label{eq:F_SSCHA}
\ee

\ni where $\mathbf{R}$ indicates a general ionic configuration and $\rho_\mathcal{H}(\mathbf{R})=\langle \mathbf{R}|e^{-\b\mathcal{H}}|\mathbf{R}\rangle/Z_\mathcal{H}$ is the probability density of finding the crystal in a generic configuration $\mathbf{R}$ ($Z_\mathcal{H}=\mathrm{tr}[e^{-\b \mathcal{H}}]$ is the partition function). By stochastically sampling configurations $\mathbf{R}$ for a given density matrix $\rho_\mathcal{H}$, and determining the gradient of the free energy with respect the harmonic parameters (centroids and force constant matrix elements), the free energy can be minimized self-consistently.
SSCHA is thus a nonperturbative method, as it does not rely on a perturbative expansion: it explores configurations with amplitudes that depend on temperature, without being limited to small displacements.

Thanks to the variational ansatz, the SSCHA quantum free energy is rigorously defined, even in out-of-equilibrium and unstable configurations like SP. In contrast, the QHA cannot determine the barrier, since the harmonic frequencies are not well defined at SP. 

The logarithm term in the harmonic term Eq.~\eqref{eq:F_harmonic} can be hard to converge for low frequencies. Separating the total free energy as in Eq.~\eqref{eq:F_SSCHA} allows to interpolate the auxiliary harmonic term up to a very fine $q$-mesh, which is fundamental to avoid oversampling the energy of the long-ranged shear modes, while the anharmonic contribution can be properly converged already with small simulation cells. So rather than calculating $F_\mathcal{H}$ directly with the  grid corresponding to the SC used in the minimization procedure, the force constant matrix can be Fourier interpolated in the usual way (building the force constant in real space from the existing $q$-grid, and then Fourier transforming back to any desired value of $\mbf{q}$), and $F_\mathcal{H}$ can be calculated using a dense mesh.

This procedure is followed both at SP and the equilibrium position, and the barrier is obtained by taking the difference of the corresponding free energies. More technical details are provided in the SI, Sec.~S1.%Sec.~\ref{sec:barrier}. 

The SSCHA frequencies are in principle auxiliary frequencies that minimize the free energy. In a purely harmonic system, such frequencies correspond to the physical frequencies. But in general, they have to be corrected with the so-called ``bubble correction'' (and also an additional higher order term in very anharmonic cases)\cite{Monacelli2021}, which we do in Sec.~\ref{sec:shear_freq} to obtain the temperature dependence of the shear frequency. Convergence with the supercell size of both the physical and auxiliary frequencies can be seen in the SI, Fig.~S7. %Fig.~\ref{fig:shearfreq_phys_exp}.

\textcolor{black}{Since the centroids of the atoms are also parameters in SSCHA, from the minimization of the free energy we additionally obtain the lattice parameters as a function of temperature, allowing us to study their thermal expansion. This is discussed in Sec.~III C.}

\section{Results}
\label{sec:results}

\subsection{Barrier}

\begin{figure}
\includegraphics[width=0.9\linewidth]{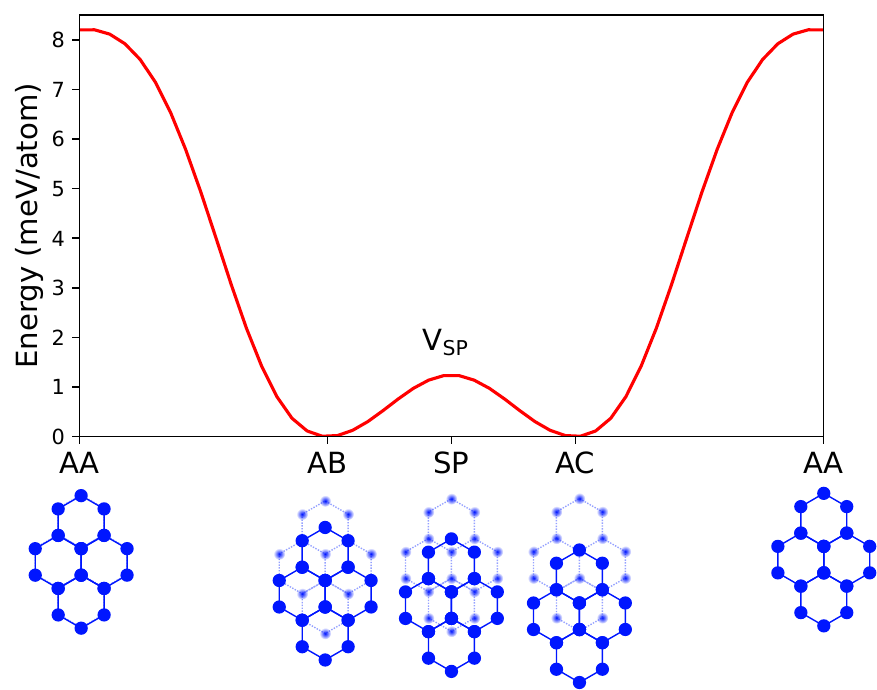}\\
\caption{Energy profile of bilayer graphene when moving one layer relative to the other along the bond direction. The lower layer is fixed in position A. Configuration AA (one layer on top of each other) is the least favorable configuration. There are two minima of the same energy, AB and AC, which correspond to the two stable configurations. To go from one minima to the other one applying shear stress, the upper layer has to overcome the barrier $V_\mathrm{SP}$, which has a value of 1.24 meV/atom relative to the minima. The diagrams below illustrate the different configurations, with the upper layer (darker blue) moving relative to the lower one in position A (lighter blue).} 
\label{fig:profile1D}
\end{figure}

We first focus on studying bilayer graphene, where the in-plane and (most importantly) out-of-plane interactions are given by the interatomic potential of Ref.~\citenum{Libbi2020} and Ref.~\citenum{Wen2018}, respectively. The out-of-plane interactions determine the barrier. The energy profile of a layer moving with respect to the other along the bond direction, shown in Fig.~\ref{fig:profile1D}, \textcolor{black}{is obtained by fixing the in-plane positions of the layers and relaxing the interlayer distance until the out-of-plane components of the atomic forces vanish,} and gives good agreement with theory and experiment (see Table~\ref{tab:energies}).

\begin{table}[h]
%\begin{tabCaption}
\caption{Comparison of the barrier $V_\mathrm{SP}$ and the shear frequency at fixed ionic positions (classical at $T=0$ K) between the interatomic potential we used in this work\cite{Libbi2020,Wen2018}, LDA, and theoretical calculations and experiments from previous works. The values are a little bit lower but close to that of adiabatic-connection fluctuation-dissipation theorem within the random phase approximation (ACFDT-RPA)\cite{Zhou2015} and experimental data\cite{Lebedeva2017}.}
{\begin{tabular}{cccc}
\hline \hline
\multicolumn{1}{l}{}  & \begin{tabular}[c]{@{}c@{}} 2 layers \\ This work\end{tabular} &  \begin{tabular}[c]{@{}c@{}}2 layers\\ LDA \end{tabular}  &\begin{tabular}[c]{@{}c@{}} Previous \\works \end{tabular} \\
\hline \\
\begin{tabular}[c]{@{}c@{}}Small barrier $V_\textrm{SP}$  \\ (meV/atom) \end{tabular}  & 1.24   & 1.58                & 1.53 (RPA) \cite{Zhou2015}           \\
\hline \\
\begin{tabular}[c]{@{}c@{}} Shear frequency \\ (cm$^{-1}$) (bulk) \end{tabular} & 37  & 42 &  42-45  \cite{Lebedeva2017} \\
\hline \hline \\
\end{tabular}} \\
\label{tab:energies}
\end{table}

The lower layer is fixed in position A, and there are two minima of the same energy, AB and AC, at a distance of a bond length $d=1.42$ \AA. For layers to slide with respect to each other, they have to overcome a barrier $V_\mathrm{SP}$. \textcolor{black}{Is is worth noting that although the profile is for bilayer graphene, some of us previously showed in first-principles calculations that using 6 layers -- 3 layers displaced relative to the other 3 -- the profile is almost identical (see SI of Ref.~\citenum{Nery2020}). Since interactions with subsequent layers are even weaker, the barrier should stay essentially unchanged with additional layers. In addition, in Ref.~\citenum{Nery2021} we obtained that energy differences between different functionals for different stacking sequences, including van-der-Waals corrections, differ by around 0.01 meV/atom. Thus, the barrier in the bulk limit should correspond closely to the bilayer one.}

\begin{figure}
\includegraphics[width=0.9\linewidth]{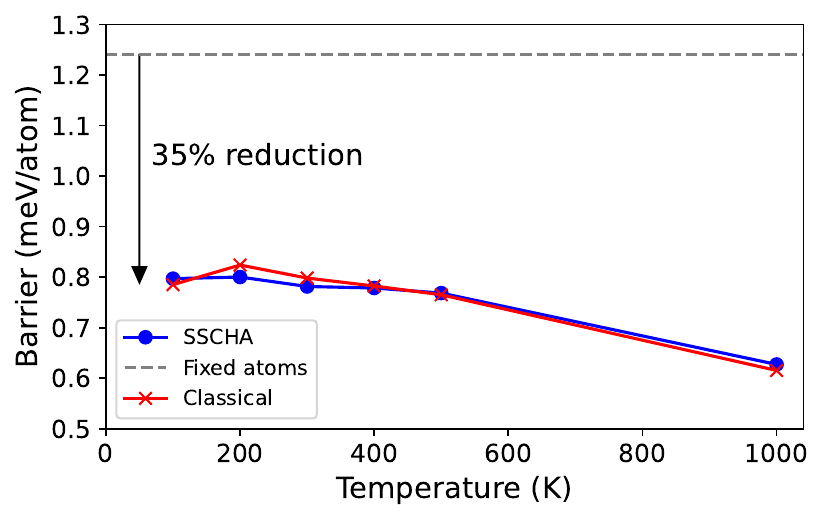}
\caption{Barrier as a function of temperature. It is reduced from 1.24 meV/atom, the value at fixed nuclei (V$_\mathrm{SP}$ in Fig.~\ref{fig:profile1D}), to about 0.8 meV/atom, corresponding to a large reduction of about 35\% due to phonons. The barrier does not change much up to about 500 K, but the value at 1000 K is about 20\% smaller relative to the value at 100 K. The lines are a guide to the eye, as in the other figures of this work. \textcolor{black}{The standard quantum SSCHA result (blue) and its classical limit (red) give almost identical results, showing quantum effects are negligible above 100~K.}}
\label{fig:barrier_interp}
\end{figure}

SP is a saddle point in the full energy landscape, so the shear mode in the perpendicular direction (along the direction of two nearby AA maxima) is stable. In order to determine temperature effects on the barrier, we use SSCHA to determine the free energy at the equilibrium position and SP, using the interpolation method described in Sec.~\ref{sec:SSCHA}. The barrier as a function of temperature can be seen in blue in Fig.~\ref{fig:barrier_interp}.

The barrier at 100 K is reduced by a substantial 35\% with respect to the value at fixed nuclei. We expect intralayer modes to contribute little to the reduction of the barrier, since they should not change much when displacing the layers. On the other hand, modes involving the relative motion of the layers should be much more affected. Indeed, intralayer modes are virtually identical, so they do not contribute to a change in the barrier. Regarding the interlayer modes, the breathing mode changes little, while the shear modes do differ significantly (see Fig.~S4).
Thus, the shear modes are responsible for the large reduction of the barrier.
 This makes sense intuitively: modes that involve a relative displacement of the layers should facilitate sliding. In the fixed ion calculation, the system is exactly at SP, but in reality, the system is also exploring surrounding configurations, which have a lower energy. \textcolor{black}{In an analogous fashion, there is a barrier in real space, since the interlayer distance at AB is lower relative to SP. At 100 K, this barrier gets similarly reduced by 38\%, from $0.029\,\text{\AA}$ to $0.018\,\text{\AA}$,
and also shows a temperature dependence similar to that of the free energy barrier (see Figs.~S5 and S6).}

Remarkably, our results show improved agreement with experiments. In a previous work\cite{Nery2020}, using density functional theory with an LDA functional (see Table 1), some of us obtained that the maximum shear stress in simulated stick-slip motion is 0.2 GPa, and we mentioned a good agreement with experimental values 0.14 GPa (Ref.~\citenum{ZeLiu2012}) --a benchmark for defect free single crystal graphite\cite{Blomquist2019}-- and 0.1 GPa (Ref.~\citenum{Dienwiebel2004}). Assuming that the maximum shear stress decreases by 35\% in line with the barrier, we now obtain 0.13 GPa, in excellent agreement with experiments\cite{Dienwiebel2004,ZeLiu2012}.

Since 100 K -- about 70 cm$^{-1}$ in wavenumber units -- is a relatively low temperature, quantum fluctuations may play a role. To assess their impact on the free energy barrier, we also performed calculations \textcolor{black}{with the classical distribution, corresponding to the classical limit $\hbar \omega / k T \rightarrow 0$. In this way, $2n_s + 1 \rightarrow 2 T/\omega$, and the quantum (Bose-Einstein) probability distribution for ionic configurations reduces to the classical Boltzmann distribution (an explicit expression for the distribution is included in the SI). We can see the SSCHA results in the classical limit in red in Fig~\ref{fig:barrier_interp}, which are virtually identical to the SSCHA quantum calculations,
 showing that quantum effects are negligible above \SI{100}{\kelvin}.}
  The reason is that shear modes responsible for the barrier reduction have very low energies, and even a modest temperature can populate such modes.
This is not a general feature, as other systems present significant differences between classical and quantum free energy barriers, as in materials formed by light hydrogen atoms. Such quantum effects have been observed at 100 K (Ref.~\citenum{Mills1994}), 250 K (Ref.~\citenum{McIntosh2013}), and even higher temperatures\cite{Poma2012}.

The size of the graphene flakes considered in our work in principle correspond to
nanometers up to possibly micrometer scales. In clean samples with micrometer sized flakes, different regions can have different stackings (AB and AC)\cite{Henck2018, Jiang2018}, and a domain wall in the metastable state separates both regions. The displacement of layers in this case is more complex and is not considered in our work. We expect shear stress to be lower in this case, since the metastable state is already present. Defects and stacking faults also lead to lower friction due to incommensurabilities, so the shear stress of 0.14 GPa corresponds to the maximum expected value \cite{Blomquist2019}.

\textcolor{black}{When it comes to displacing many layers in multilayer graphene or graphite, the change of the barrier could in principle differ, since it depends on the phonon frequencies (and interlayer modes depend noticeably on the number of layers). 
For example, graphite exhibits a larger shear frequency at equilibrium, since a layer experiences primarily the restoring force of two layers instead of one (see the next subsection for more details). However, layers are typically not simultaneously in unstable positions relative to both nearest layers, so, as noted earlier, the barrier remains approximately unchanged in bulk. Since the frequency at AB increases while remaining similar at SP, the barrier reduction in bulk might be even greater than in the bilayer case. Other variations may arise from the choice of the interlayer interatomic potential: as shown in Table I, the RPA barrier is higher, indicating that the curvature around SP could be larger. Thus, vibrations around SP with more accurate interlayer potentials might yield a larger absolute barrier reduction, although the relative change may still be similar.}

In other materials, the anharmonicities associated with the unstable modes at a saddle point should also contribute negatively to the free energy. However, the overall change of the barrier will depend on differences in the entire SSCHA phonon spectra at the minima and saddle point.

Regarding the evolution of the barrier as the temperature increases, there is a weak dependence up to about 500 K, since the barrier remains close to around 0.8 meV/atom. However, when temperature increases up to 1000 K, the barrier is further reduced by 20\% relative to the value at 100 K. Thus, high temperatures should facilitate sliding between layers. In particular, it could aid the production of ABC-stacked multilayer graphene from its more common AB-stacked version through shear stress, as mentioned in the Introduction. More in general, high temperatures could also help engineer specific stacking arrangements in other layered materials.

\vspace{-0.2cm}
\subsection{Shear frequency}
\label{sec:shear_freq}  
\vspace{-0.2cm}

\begin{figure}[h]
\centering
\includegraphics[width=0.95\linewidth]{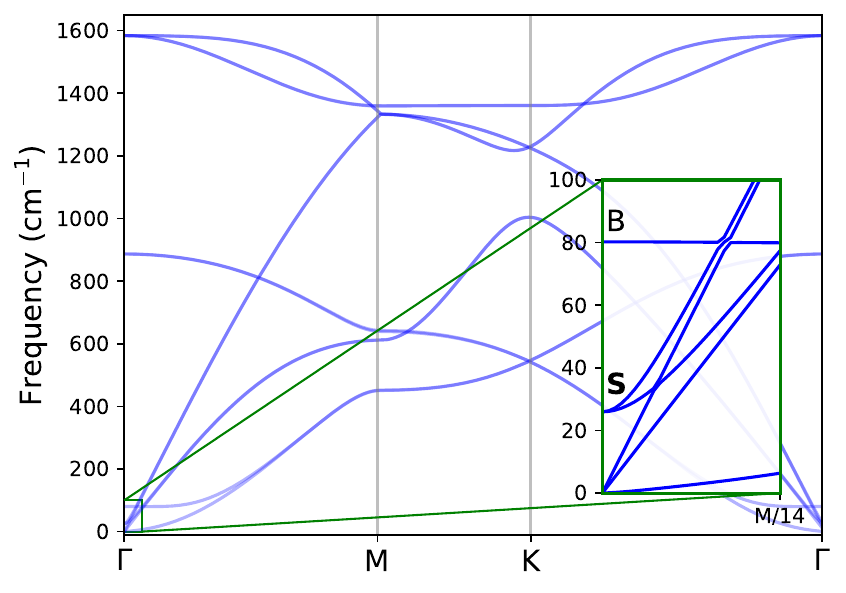}\\
\caption{Bilayer graphene phonon dispersion. The inset shows a zoom in close to $\Gamma$ to better visualize the doubly degenerate shear modes \textbf{S} at $\Gamma$, and the breathing mode B. Due to the high speed velocity in graphene, the region of the shear mode in the BZ is small (a fraction of $\Gamma$M), and is overrepresented in small supercell calculations. The breathing mode, on the other hand, is flat, so nearby sampling points are not necessary to converge the free energy.}
\label{fig:phonon_dispersion}
\end{figure}

At the equilibrium position, one of the vibrational modes at $\Gamma$ is the shear mode in both directions of the plane (doubly degenerate, see Fig.~\ref{fig:phonon_dispersion}).
Its temperature dependence can be determined using both SSCHA and QHA. In the QHA, the TE is obtained by determining the lattice parameters that minimize the free energy at each temperature. 

The third out of phase mode of the layers is the breathing mode, in the out-of-plane direction. A discussion of its temperature dependence can be found in the SI.

\begin{figure}
\includegraphics[width=0.9\linewidth]{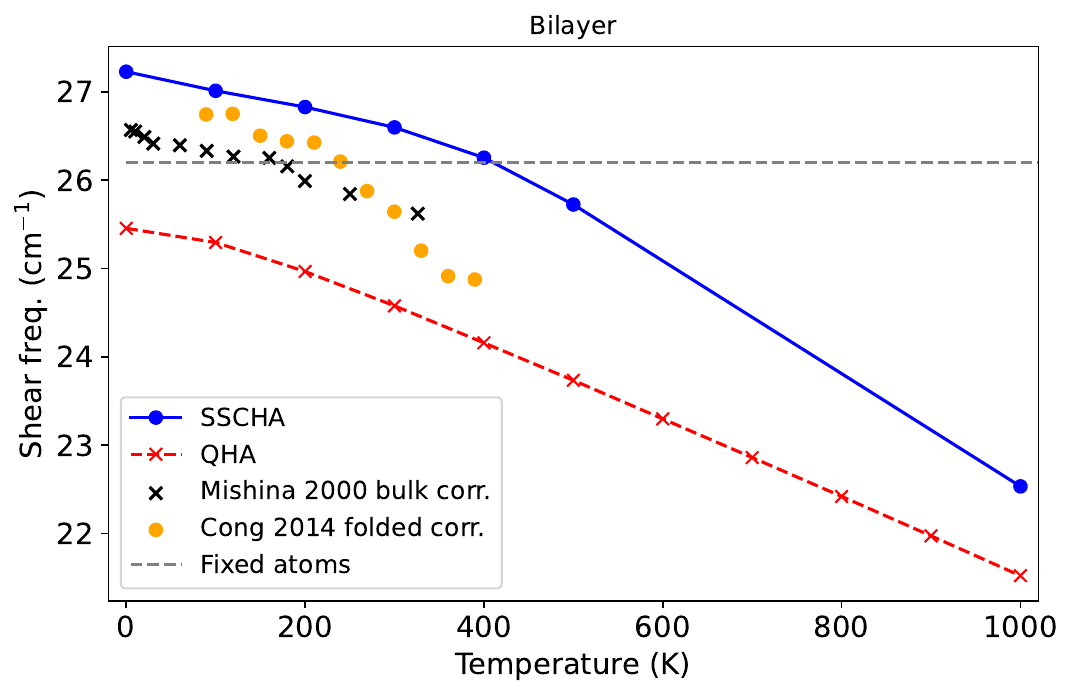}\\
\caption{Shear frequency as a function of temperature using SSCHA (blue circles) and QHA (red crosses). There is no experimental data for the temperature dependence of the shear frequency as a function of temperature for bilayer graphene. Experimental values correspond to bulk (blue crosses) and folded 2+2 graphene (orange circles), which are corrected by the factor of a nearest neighbor model\cite{He2016} to compare to the bilayer case (and rigidly shifted by -5 cm$^{-1}$ to facilitate the visual comparison). The agreement with the temperature dependence of the bulk values is good.}
\label{fig:freq}
\end{figure}

The temperature dependence of the shear frequency can be seen in Fig.~\ref{fig:freq}. 
The quantum effects evaluated within the SSCHA harden the shear frequency energy at low temperatures compared with the classical value at 0~K (dashed line).
On the other hand, the zero-point renormalization in QHA softens the shear frequency.
 The temperature dependence of the shear mode is similar to that of the barrier, varying by about 20\% from 0 to 1000 K, and having a larger dependence at temperatures above 500 K. 

\textcolor{black}{Regarding the experimental data, in Ref.~\citenum{Cong2014} the measurements are done in 2+2 folded graphene (4 layers), and in Ref.~\citenum{Mishina2000} they are done in bulk. To compare with our bilayer calculations, we rely on a nearest layer model which has simple conversion factors between different number of layers\cite{Tan2012}. Since interlayer interactions in graphite are known to be weak, this should provide a good approximation. Indeed, there are multiple references that have measured the shear frequency for different number of layers and obtained that the model describes the trend very well, like in multilayer graphene \cite{Tan2012} and NbSe$_2$ \cite{He2016}. The bulk case can be readily understood: while in the bilayer case the restitutive lateral force is exerted only by the other layer, in the bulk case each layer is surrounded by two of them, so the force is doubled. Since the interatomic force constant matrix is proportional to the square of the phonon frequencies, the shear frequency in bilayer is $\sqrt{2}$ smaller than the bulk frequency in this model. 
Using these factors\cite{Tan2012}, we obtain the adjusted experimental values of Fig.~\ref{fig:freq}.}

 The temperature dependence of both SSCHA and QHA curves is similar to the experimental adjusted bulk values. Folded graphene values have a steeper dependence but are likely unsuitable for comparison, because layers fold with different rotational angles, while layers are aligned in standard graphite. The shear frequency has also been measured in other materials like bilayer NbSe$_2$ in Ref.~\citenum{He2016}, or bulk $h$-BN in Ref.~\citenum{Cusco2019}, which show a similar temperature dependence (adjusting the bulk value by a $\sqrt{2}$ factor) of roughly -0.5 cm$^{-1}$/100 K. 

\subsection{Thermal expansion}

\begin{figure}[!ht]
\subfigure[]{\includegraphics[width=0.85\linewidth]{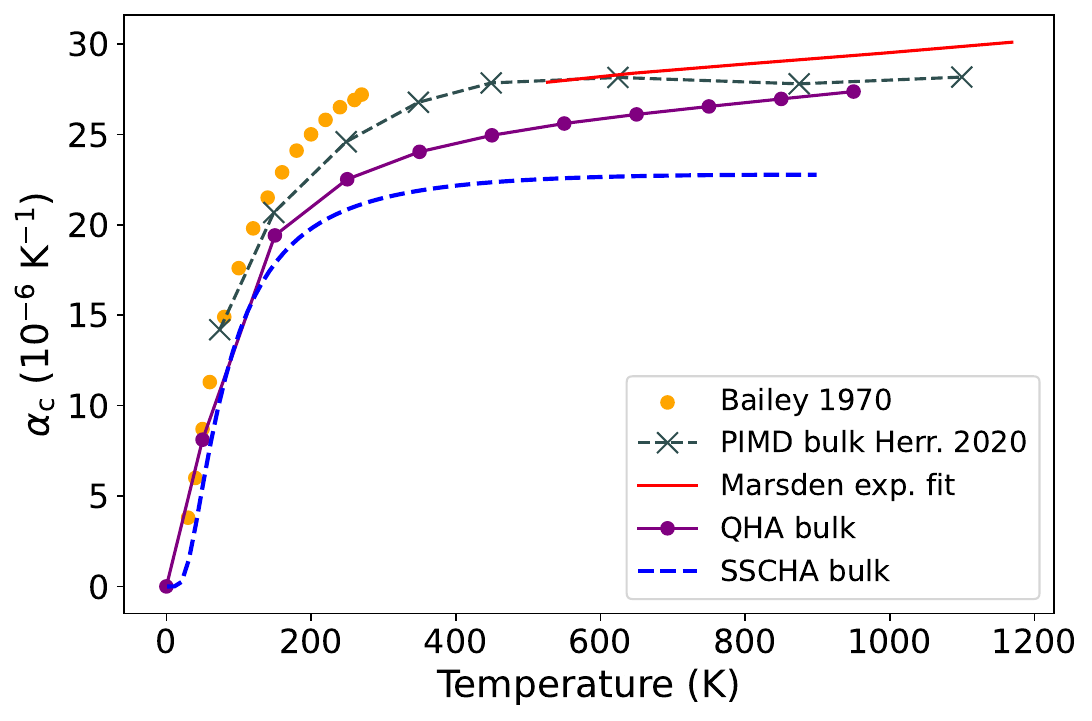}}
\subfigure[]{\includegraphics[width=0.9\linewidth]{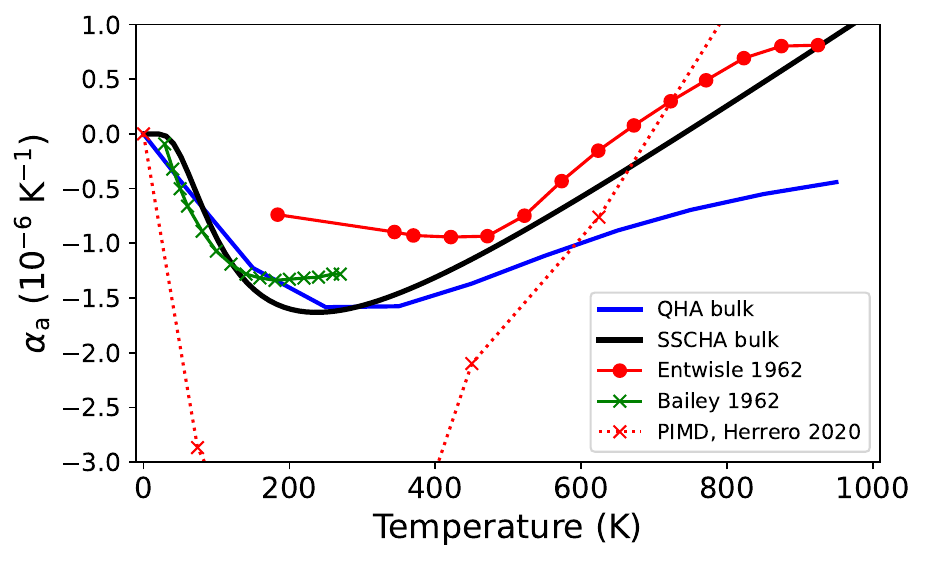}}
\caption{(a) Graphite out-of-plane coefficient of TE as a function of temperature using SSCHA and QHA. PIMD calculations\cite{Herrero2020} and experimental data \cite{Bailey1970} are also included. PIMD overall agrees well, while QHA and SSCHA (with the potential of Ref.~\citenum{Wen2018}) underestimates the TE. (b)
Graphite in-plane coefficient of TE using SSCHA and QHA. QHA again underestimates the dependence at low temperatures, and also at high temperatures. On the other hand, PIMD gives a dependence that is surprisingly very far from experimental values. SSCHA gives excellent results both at low and high temperatures, considering also the disagreement within experimental data.}
\label{fig:thermal_expansion}
\end{figure}

To further check the adequacy of our method (the use of SSCHA with the in-plane\cite{Libbi2020} and out-of-plane\cite{Wen2018} interatomic potentials), we look at the TE in the previous calculations and also in bulk graphite, and compare to experiments. The TE coefficients are given by 
 $\a_\mathrm{a}(T) =\tf{1}{a} \tf{da}{dT}$ and $\a_\mathrm{c}(T)=\tf{1}{c}\tf{dc}{dT}$, with $a$ and $c$ the in-plane and out-of-plane lattice parameters, respectively.

In Fig.~\ref{fig:thermal_expansion}, we show the TE coefficient calculated with SSCHA and QHA in bulk, and compare with experimental data\cite{Bailey1970,Marsden2018} and PIMD calculations~\cite{Herrero2020}. 
SSCHA results underestimate the TE \textcolor{black}{, which we attribute to the interlayer potential\cite{Wen2018}. 
This potential is fitted to energies and forces that use the many-body dispersion method in conjunction with a PBE functional, as it reproduces binding energies and interlayer distances reasonably well. However, it yields an elastic modulus along the c-axis about 15\% lower than experimental values (see more details within Ref.~\citenum{Wen2018}). Unsurprisingly, Ref.\citenum{Herrero2020} obtains better results with the LCBOPII potential\cite{LCBOPII} -- although the TE remains somewhat below experiments\cite{Bailey1970,Marsden2018} -- since it uses the experimental out of plane compressibility as a fitting parameter\cite{Karssemeijer2011}. The bilayer TE is also similarly underestimated (see SI). Thus, the variation of the potential around the interlayer equilibrium distance seems to be the main factor influencing the TE, as opposed to the weaker long-range interactions involving subsequent layers. It is also possible that anharmonicities, captured more accurately by PIMD relative to SSCHA, play a role at high temperatures.
In any case, for the purpose of studying the barrier -- the focus of our work -- the problem with the LCBOPII potential is that it includes no registry effects, making the barrier essentially non-existent.} 
 
\textcolor{black}{While the out-of-plane TE we obtained is not optimal,} in-plane SSCHA calculations give excellent results. QHA works well at low temperatures but departs from experiments at higher temperatures.
\textcolor{black}{The negative TE in graphite and other layered systems is well-known and arises from flexural (out-of-plane) modes \cite{Mounet2005}. We briefly summarize the mechanism here. In flexural modes, atoms are displaced out-of-plane, reducing the average projected in-plane distance, even though the 3D bond lengths remain essentially unchanged. This can also be understood intuitively via the “membrane effect”: when a layer is stretched, the out-of-plane amplitude decreases and the frequency increases, which corresponds to a negative Grüneisen parameter. See, for example, Ref.~\citenum{Mounet2005} for more details.}
We conducted a careful convergence of these results, reported in the SI, Sec.~S3. 
PIMD calculations of Ref.~\citenum{Herrero2020}, although they work very well out-of-plane, greatly overestimate in-plane TE, presumably due to the inadequacy of their interatomic potential. 
 It would be interesting in future work to compare SSCHA and PIMD using the same potentials. 

\section{Conclusions}
\label{sec:conclusions}
To conclude, we demonstrated that ionic vibrations %anharmonicities 
are fundamental to determining properties of layered materials that depend on free energy barriers. Using the self-consistent harmonic approximation, which can calculate the free energy at unstable configurations, and a novel interpolation method, we unveiled the suppression of the barrier separating stable configurations in bilayer graphene. Phonon vibrations reduce the free energy barrier by a significant 35\%, and gets further suppressed at high temperatures, so barriers at fixed ions cannot be trusted. Now, correcting the shear stress of 0.2 GPa reported in our previous work, we arrive at a value of 0.13 GPa, in excellent agreement with experiments. 
The shear modes, which occupy a small region of phase space, have to be adequately sampled to correctly determine the free energy barrier as well as the thermal expansion. Although our approach underestimates the out-of-plane thermal expansion compared with experimental values, in-plane agreement with experiments is excellent. We also obtained that the temperature dependence of the barrier is similar to that of the shear frequency, which agrees well with existing measurements. \\

\section*{Data availability}
\vspace{-0.3cm}
\ni Additional data supporting this article has been included as part of the Supplementary Information. The SSCHA\cite{Monacelli2021} code used in this article is publicly available at https://sscha.eu/. 

\vspace{-0.15cm}
\section*{Conflicts of interest}
\vspace{-0.3cm}
\ni There are no conflicts to declare.

\vspace{-0.15cm}
\section*{Acknowledgments}
\vspace{-0.3cm}
\ni This work has received funding from the European Union’s Horizon 2020 research and innovation
programme under Grant Agreement No. 881603, and from the MORE-TEM ERC-SYN project, Grant Agreement No. 951215. J.P.N. is currently supported by the European Union under a Marie
Sk\l{}odowska-Curie Individual Fellowship, Project GreenNP No. 101151380.

\bibliography{bibliography}

\end{document}